\documentclass[11pt]{article}

\usepackage{amsmath,amsfonts,amssymb,amsthm,epsfig, graphicx, hyperref}
\usepackage[dvipsnames,table,dvipsnames*, svgnames*, hyperref]{xcolor}
\usepackage{algorithm}
\usepackage{setspace}
\usepackage{natbib}
\usepackage{empheq}
\usepackage{caption}
\usepackage{subcaption}
\usepackage{babel,blindtext}
\usepackage{mwe}
\usepackage{float}
\usepackage{mathrsfs,enumitem,Notations}
\usepackage{enumerate}
\usepackage{booktabs}
\DeclareMathSizes{12}{12}{5}{3}
\usepackage{multirow}%
\usepackage[title]{appendix}%
\usepackage{textcomp}%
\usepackage{manyfoot}%
\usepackage{booktabs}%
\usepackage{algpseudocode}%
\usepackage{listings}%
\usepackage{empheq}
\usepackage{lscape}
\usepackage{diagbox}
\usepackage{tabularx}
\usepackage{rotating}

\def\bX{\mathbf{X}}
\def\bbeta{\boldsymbol{\beta}}
\def\balpha{\boldsymbol{\alpha}}

\def\bdelta{\boldsymbol{\delta}}
\newcommand\independent{\protect\mathpalette{\protect\independenT}{\perp}}
\def\independenT#1#2{\mathrel{\rlap{$#1#2$}\mkern2mu{#1#2}}}

\newcommand{\blind}{1}

\begin{document}

\def\spacingset#1{\renewcommand{\baselinestretch}%
{#1}\small\normalsize} \spacingset{1}


\if1\blind
{
  \title{\bf A Transfer Learning Causal Approach to Evaluate Racial/Ethnic and Geographic Variation in Outcomes Following Congenital Heart Surgery}
  \author{Larry Han$^1$, Yi Zhang$^2$, Meena Nathan$^3$, John E. Mayer, Jr.$^3$, \\ Sara K. Pasquali$^4$, Katya Zelevinsky$^5$, Rui Duan$^6$, Sharon-Lise T. Normand$^5$ \\  
  \hspace{.2cm}\\
    1 Department of Health Sciences, Northeastern University\\
    2 Department of Statistics, Harvard University \\
    3 Department of Cardiac Surgery, Boston Children's Hospital \\
    4 Division of Cardiology, Department of Pediatrics, \\ University of Michigan C.S. Mott Children's Hospital \\
    5 Department of Health Care Policy, Harvard Medical School \\
    6 Department of Biostatistics, Harvard T.H. Chan School of Public Health}
  \maketitle
} \fi

\if0\blind
{
  \bigskip
  \bigskip
  \bigskip
  \begin{center}
    {\LARGE\bf }
\end{center}
  \medskip
} \fi

\bigskip
\begin{abstract}
    Congenital heart defects (CHD) are the most prevalent birth defects in the United States and surgical outcomes vary considerably across the country. The outcomes of treatment for CHD differ for specific patient subgroups, with non-Hispanic Black and Hispanic populations experiencing higher rates of mortality and morbidity. A valid comparison of outcomes within racial/ethnic subgroups is difficult given large differences in case-mix and small subgroup sizes. We propose a causal inference framework for outcome assessment and leverage advances in transfer learning to incorporate data from both target and source populations to help estimate causal effects while accounting for different sources of risk factor and outcome differences across populations. Using the Society of Thoracic Surgeons' Congenital Heart Surgery Database (STS-CHSD), we focus on a national cohort of patients undergoing the Norwood operation from 2016-2022 to assess operative mortality and morbidity outcomes across U.S. geographic regions by race/ethnicity. We find racial and ethnic outcome differences after controlling for potential confounding factors. While geography does not have a causal effect on outcomes for non-Hispanic Caucasian patients, non-Hispanic Black patients experience wide variability in outcomes with estimated 30-day mortality ranging from $5.9\%$ (standard error $2.2\%$) to $21.6\%$ ($4.4\%$) across U.S. regions.
\end{abstract}

\noindent%
\vfill

\newpage
\spacingset{1.25} 

\section{Introduction}\label{sec1}

Congenital heart defects (CHD) are the most prevalent birth defects in the U.S., impacting approximately one in every $100$ births. Annually, there are more than $40{,}000$ pediatric congenital heart surgeries (CHS), accounting for significant resource utilization across the nation's children's hospitals \citep{pasquali2012variation, pasquali2016congenital, pasquali2019development}. Although surgical outcomes have improved over time, certain defects remain associated with high morbidity and mortality, upwards of $10-15\%$ in some cases \citep{kumar2023society}.

Wide variability in outcomes has been reported at both the patient-level (associated with certain risk factors such as age, weight, syndromes, specific cardiac diagnoses, etc.) as well as at the broader level across centers and regions \citep{normand2022, knowles2019ethnic, udine2021geographical, schwartz2023social}. However, gaps remain in our ability to focus on specific subgroups to best understand risk due to the wide heterogeneity in types of heart disease (e.g., hundreds of different diagnoses), resultant small sample sizes in any given group, and varying impact of different risk factors in and across diagnostic and procedural categories \citep{jacobs2019refining, pasquali2019development, o2019development, bush2018clinical, bush2021pulmonary}.

For example, it is known that there is substantial variability in outcomes both across geographic locations \citep{guadagnoli2001impact, udine2021geographical, millen2023effect, schwartz2023social} as well as by patient race, ethnicity, and socioeconomic factors \citep{haider2013racial,  richardson2021racial, milam2024racial, krishnan2021impact, schwartz2023social}. Additionally, lower neighborhood socioeconomic status is associated with worse survival outcomes following the Norwood operation \citep{bucholz2018neighborhood, sengupta2023impact}. However, how these factors relate and interact among subgroups of patients is not clear. This information is critical in informing interventions to reduce variation and improve outcomes. Furthermore, prior studies are correlative and the findings may not generalize to out-of-sample studies where heterogeneity in the data distribution and varying local practices affect model performance.
To our knowledge, no prior studies have examined racial and ethnic geographic variation in post-operative outcomes for specific CHS from a causal inference framework.

We propose an innovative causal inference framework and methodology to address these challenges, and in this study, we focus on understanding the relationship between geographic region and race and ethnicity on outcomes among patients undergoing the Norwood operation. This operation is the first in a series of three staged operations performed for patients with hypoplastic left heart syndrome (HLHS) and other types of single ventricle heart disease, is typically performed in the first few weeks of life, and remains associated with significant morbidity and mortality in the current era \citep{norwood1983physiologic, sano2003right, galantowicz2008hybrid, sperotto2023transcatheter}. Our proposed method defines the \emph{target} population, that is the population of interest, to be a specific racial and ethnic group undergoing a Norwood procedure in a particular geographic region. 
To overcome the impact of small sample sizes, we leverage advancements in transfer learning \citep{li2020transfer} to incorporate data from both the target population and other potentially similar populations, denoted \emph{source} populations, to improve estimation efficiency. Using a national, contemporary cohort of patients undergoing Norwood procedures 
in the U.S., our analysis answers the question: Would we expect different outcomes had patients undergone their Norwood procedures in a different region of the country, and if so, do these differences differ across racial and ethnic groups? 

Our study finds that geography does not have a causal effect on outcomes for non-Hispanic Caucasian patients. Post-surgical outcomes vary little -- estimated 30-day mortality ranges from $10.6\%$ (standard error $1.6\%$) to $13.2\%$ ($1.8\%$) and 30-day morbidity ranges from $24.3\%$ ($1.7\%$) to $31.1\%$ ($2.5\%$)) across geographic regions.  Conversely, non-Hispanic Black and Hispanic patients experienced greater geographic outcome variability, with the most pronounced differences observed among non-Hispanic Black patients.  For these patients, the estimated 30-day mortality ranges from $5.9\%$ ($2.2\%$) to $21.6\%$ ($4.4\%)$ and 30-day morbidity from $23.0\%$ ($4.0\%$) to $43.3\%$ ($9.1\%$)) across geographic regions. Non-Hispanic Black patients were predicted to experience higher mortality and morbidity rates in the West or Midwest regions compared to the South or Northeast regions. These findings highlight that post-surgical outcomes differ across geographic regions for specific racial and ethnic groups and underscore the need to better understand the reasons for this variability.

\section{Methods}
\subsection{Causal inference framework}
 
To estimate regional effects in a population of interest (referred to as the target population), a straightforward way is to use data only from the target population. This target-only strategy is undesirable and potentially infeasible when the target population has a limited sample size as it can lead to poor model accuracy and low statistical efficiency. Another strategy is to combine data across all populations to fit the required nuisance models. However, this strategy is subject to potential bias when the target and the non-target populations (referred to as source populations) have different underlying outcome and risk factor distributions. Three sources of distribution shift between target and source populations are possible. These include covariate shift (e.g., the target population may have more socioeconomically disadvantaged patients), propensity score model shift (e.g., patients with the same covariate profile in the target and source populations may not have the same probability of residing in a specific region of the country), and outcome regression model shift (e.g., patients with the same covariate profile in the target and source populations may be expected to have different outcomes). Transfer learning methods improve model fitting in the target population and account for potential distributional shifts. In this paper, we define the target population to be one of the 12 combinations of racial and ethnic groups (three) and geographic regions (four) and the source populations to be the other eleven groups. 

We estimate the mean potential outcomes for each target population, as well as causal differences if a certain racial and ethnic group treated in a specific geographical region were to receive surgery in another geographical region, contrary to fact. To estimate this causal effect, we fit an outcome regression model within each geographical region using a logistic regression model incorporating all of the measured covariates. We estimate a density ratio to balance the covariate distribution across geographical regions so that any observed disparities in outcomes are not due to measured confounding. When studying a target population, observations from source populations are incorporated using a transfer learning method \cite{li2020transfer} to leverage the shared similarities between the outcome regression models across groups.


\subsection{Transfer learning approach for quality measurement}
For each individual $i$, let $Y_{i} \in \mathbb{R}$ denote an observed outcome, which can be continuous or discrete and $X_{i} \in \mathbb{R}^{p}$ represents the $p$-dimensional baseline covariates. Suppose there are $M$ geographical regions and $K$ {race and ethnicity} groups. Let $T$ denote region and $R$ denote {race and ethnicity} group, and each individual $i$ belongs to exactly one $(R,T)$ stratum. Define $n_{k,m}$ as the sample size within each stratum $(R=k,T=m)$ and $n = \sum_{k,m} n_{k,m}$ as the total sample size. For simplicity of presentation, we suppress subscript $i$ to state general assumptions and conclusions. We define the potential outcomes as $Y(m)$, which is a function of the multi-valued region $m\in\{1,\ldots,M\}$. The target parameter of interest is the target average treatment effect of the treated (TATT)
\begin{equation}
    \tau_{k,m_1,m_2}:=\mathbb{E}\left[Y(m_1)-Y(m_2)\mid R=k,T=m_2\right] 
\end{equation}
for regions $m_1,m_2\in\{1,\ldots,M\}$ and race-ethnicity group $k\in\{1,\ldots,K\}$. In our setting, the TATT is the target average region effect for individuals receiving care in a particular region, i.e., the difference in outcomes had patients of a specific {race and ethnicity} received care in a different region, given that they received care in a particular region.

For nonparametric identification of the TATT, we
will adopt the following causal assumptions \citep{robins1986new}:
\begin{description} 
    \item[Assumption 1: Identification of TATT]
    \item[(a) Consistency:] $Y(T) \equiv Y$ 
    \item[(b) Positivity:] $\mathbb{P}(T = m \mid \boldsymbol{X}, R) > \epsilon,$
   w.p. 1, for some $\epsilon > 0$ and  $m \in \{1,...,M\}$
    \item[(c) No unmeasured confounding:] $Y{(m)} \independent T \mid \boldsymbol{X}, R
    \text{ for } m \in \{1,...,M\}$ 
\end{description}

Consistency says that we observe the counterfactual corresponding to the observed region level, positivity says that each region is possible in all strata defined by $(\boldsymbol{X}, R)$, and no unmeasured confounding assumes that within levels of 
$(\boldsymbol{X}, R)$, region $T$ is as good as randomized.

Given these assumptions, the estimand $\tau_{k,m_1,m_2}$ can be nonparametically identified as 
\begin{equation}
    \tau_{k,m_1,m_2}=\mathbb{E}\left[\mathbb{E}\left[Y\mid \boldsymbol{X},R=k,T=m_1\right]\mid R=k,T=m_2\right]-\mathbb{E}\left[Y\mid R=k,T=m_2\right]
\end{equation}
Therefore, we propose a doubly-robust estimator for estimating $\tau_{k,m_1,m_2}$: 
{\small
\begin{equation}\label{equ:DR:ATT}
\begin{aligned}
    \hat{\tau}_{k,m_1,m_2}=& \frac{1}{n_{k,m_2}}\sum_{i=1}^n  \left[ \mathbf{1}(R_i=k,T_i=m_2){\mu}_{k,m_1}(X_i\right.)\\
    &\left.+ \mathbf{1}(R_i=k,T_i=m_1)\frac{\mathbb{P}(T_i=m_2\mid X_i,R_i=k)}{\mathbb{P}(T_i=m_1\mid X_i,R_i=k)}(Y_i-{\mu}_{k,m_1}(X_i))\right]\\
    &-\frac{1}{n_{k,m_2}}\sum_{i=1}^n \mathbf{1}(R_i=k,T_i=m_2)Y_i,
\end{aligned}
\end{equation}
}

\noindent
where ${\mu}_{k,m_1}(X) = {\mathbb{E}}\left[Y\mid X,R=k,T=m_1\right]$ is the conditional mean outcome for the target population $k$ in the comparison region $m_1$. Because $\hat{\tau}_{k,m_1,m_2}$ in Equation~\eqref{equ:DR:ATT} involves unknown nuisance functions, we fit a multinomial regression for the region selection propensity $\mathbb{P}(T_i=m\mid X_i,R_i=k),\; m\in\{1,\dots,M\}$ and propose a transfer learning approach (Algorithm~\ref{alg:binary}) to learn the conditional outcome expectation ${\mu}_{k,m_1}(\cdot)$. Specifically, we leverage the similarities in the conditional outcome model across different populations within the same region, which improves the accuracy of nuisance estimates, as opposed to the standard approach using solely data from the target population.

\subsection{Algorithm}
For each given region $m$ and race and ethnicity group $k$, we assume a  parametric model for the conditional expectation of outcome ${\mu}_{k,m}(X):={b}_{k,m}(X;\boldsymbol{\beta}_{k,m})$, where $\boldsymbol{\beta}_{k,m}$ is a finite-dimensional parameter that may vary across sites and populations.
Given the form of our proposed estimator in Equation~\eqref{equ:DR:ATT}, we now specifically focus on the estimation of ${\mu}_{k,m_1}$, i.e., the outcome model in the comparison region $m_1$ for a given race and ethnic group $k$.
Details on each step of our algorithm can be found in Algorithm \ref{alg:1}. Briefly, after defining geographic and racial and ethnic strata (Step 1), we first learn the outcome model coefficients from each of the source populations (Step 2), and use them to ``jumpstart" the model fitting in the target population. Specifically, we adjust for the difference in outcome coefficients from the target population by offsetting these estimated source coefficients in model fitting using target data (Step 3).
When the source models are substantially different from the target model, the learned estimators may not be better than a target-only estimator (Step 4) obtained using only the target data.
Thus, to prevent negative transfer learning, we perform an aggregation step using some additional validation data from the target population to optimally combine TL-based source estimators with the target-only estimator, yielding the final parameter estimator $\boldsymbol{\beta}_{k,m_1}$ (Step 5) and the model estimate $\hat{\mu}_{k,m_1}(\cdot)$ (Step 6).

\begin{algorithm}[H]
    \caption{Transfer learning algorithm (Target parameter: $\boldsymbol{\beta}_{k,m_1}$) }\label{alg:binary}
    \begin{algorithmic}[1]
   \State \textbf{Data: $\{(Y_i,X_i,T_i=m_1,R_i=k')\}_{k'=1}^K$ } 
   \State{%
    For each source population $k'\neq k$, fit a population-specific regression model using $\{(Y_i,X_i,T_i=m_1,R_i=k')\}$,
    \begin{equation}
       \hat{\boldsymbol{\beta}}_{k',m_1}=\underset{\boldsymbol{b} \in \mathbb{R}^p}{\arg \min }\left\{\frac{1}{n_{k',m_1}}  L^{(k',m_1)}(\boldsymbol{b})+\lambda^{(k',m_1)}\|\boldsymbol{b}\|_2\right\}.
    \end{equation}
   where $L$ is a negative log-likelihood function for the outcome model, and $\{\lambda^{(k',m_1)}\}_{k'=1 }^K$ and $\lambda_\delta$ (used below) are tuning parameters obtained via cross-validation. 
  }
\State Adjust for differences using target population data $\{(Y_i,X_i,T_i=m_1,R_i=k)\}$. For each source population $k'\neq k$, calculate the difference in outcome coefficients from the target population $ \hat{\boldsymbol{\delta}}_{k', m_1}$,
 \begin{equation}
\hat{\boldsymbol{\delta}}_{k', m_1}=\underset{\boldsymbol{b} \in \mathbb{R}^p}{\arg \min }\left\{\frac{1}{n_{k,m_1}} L^{(k,m_1)}\left(\hat{\boldsymbol{\beta}}_{k', m_1}+\boldsymbol{b}\right)+\lambda_\delta\|\boldsymbol{b}\|_2\right\}.
  \end{equation}
   Obtain $\hat{\boldsymbol{\beta}}_{k',m_1}^{\text(tar)}=\hat{\boldsymbol{\beta}}_{k',m_1}+\hat{\boldsymbol{\delta}}_{k',m_1}$. 
\State Fit a target-only estimator using $\{(Y_i,X_i,T_i=m_1,R_i=k)\}$,
\begin{equation}
       \hat{\boldsymbol{\beta}}_{k,m_1}=\underset{\boldsymbol{b} \in \mathbb{R}^p}{\arg \min }\left\{\frac{1}{n_{k,m_1}}  L^{(k,m_1)}(\boldsymbol{b})+\lambda^{(k,m_1)}\|\boldsymbol{b}\|_2\right\}.
    \end{equation}
\State With $\hat{\boldsymbol{\beta}}_{k,m_1}$ and $\{\hat{\boldsymbol{\beta}}_{k',m_1}^{\text(tar)}\}_{k'\neq k}$, we perform aggregation using some additional validation data from the target population, $\{(Y_i,X_i,T_i=m_1,R_i=k)\}_{i=1}^{\stackrel{\circ}{n}}$. Define $\widehat{\boldsymbol{B}}=\left(\{\hat{\boldsymbol{\beta}}_{k',m_1}^{\text(tar)}\}_{k'\neq k}, \hat{\boldsymbol{\beta}}_{k,m_1}\right) \in \mathbb{R}^{p \times K}$.
Compute the aggregation weights
\begin{equation}
    \hat{\boldsymbol{\eta}}=\underset{\eta \in \mathbb{R}^{K},\eta_i\geq0,\|\eta\|_1=1}{\arg \min }\left\{\sum_{i=1}^{\stackrel{\circ}{n}} Y_i \cdot X_i^{\top} \widehat{\boldsymbol{B}} \boldsymbol{\eta}-\psi\left(X_i^{\top} \widehat{\boldsymbol{B}} \boldsymbol{\eta}\right)\right\},
\end{equation}
where $\psi(\cdot)$ is a function uniquely determined by the link function in the parametric outcome model.
Compute the aggregated estimator as $\hat{\boldsymbol{\beta}}_{k,m_1}^{\text{(agg)}}=\widehat{\boldsymbol{B}} \hat{\boldsymbol{\eta}}$. 

\State Use  $\hat{\boldsymbol{\beta}}_{k,m_1}^{\text{(agg)}}$ to construct $\hat{\mu}_{k,m_1}(\cdot)$ estimates.
    \end{algorithmic}
\label{alg:1}
\end{algorithm}

\clearpage

\section{Simulation study}

To examine the performance of our estimator in finite sample sizes, we conducted a simulation study. We set $K=4$ populations and $M=5$ different regions.
The total sample size across all populations and sites is $10{,}000$, with the proportions in the four populations being $(5\%,15\%,20\%,60\%)$ to be representative of racial and ethnic distributions, leading to a sample size of  $(n_k)_K=(500, 1500, 2000, 6000)$. 

For each population $k$, we generate $\bX=\left(X_1, \ldots, X_p\right)^{\mathrm{T}}$ from multivariate normal distribution with means $\mu_k$ and covariance $\operatorname{cov}\left(X_{j_1}, X_{j_2}\right)=2^{-\left|j_1-j_2\right|}$ for $1\leq j_1,j_2\leq p$ \citep{xu2022high}. We set the dimension of $\bX$ to be $p = 20$ and truncate $\bX$ between $-2$ and $2$.

We generate region indicator $T$ given $\bX$ and $R=k$ from a categorical distribution with
$\log \{\mathbb{P}(T=m \mid \bX,R=k) / \mathbb{P}(T=1 \mid \bX,R=k)\}= \bX^\top \balpha_{k,m} $ for $m\in\{2,\ldots,M\}$ and $k\in\{1,\ldots,K\}$, where each entry of  $\balpha_{k,m}$ is generated from $\text{Unif}(-0.2,0.2)$.
The binary outcome $Y$ given $\bX$, $R=k$ and $T=m$ is generated from a binomial distribution with probability of success $1/(1+\exp(-\bX^\top \bbeta_{k,m}))$, where each coefficient $\bbeta_{k,m}$ is generated according to $\bbeta_{k,m}=\underbrace{(0.6,\ldots, -0.4)}_{\text{equally-spaced decrements}}+\bdelta_{k,m}$. Here, $\bdelta_{k,m}$ represents the difference in coefficients across populations and sites. $\bdelta_{k,m}$ has $s=5$ non-zero entries, with each non-zero entry generated from a uniform distribution over $\{0.1,0.2\}$ and assigned a positive sign with probability $0.5$.

We compare our proposed transfer learning approach with the target-only method on the most underrepresented population group $R=1$. 
We focus on the population-level TATT $\mathbb{E}\left[Y(m_1)-Y(m_2)\mid R=1,T=m_2\right]$.
The results are reported in terms of bias, RMSE and coverage probability of the 95\% CI based on 1,000 simulation runs. Our results indicate that our proposed estimator achieves lower bias (Table \ref{tab:bias}), lower RMSE (Table \ref{tab:rmse}), and better coverage probability of the $95\%$ CI (Table \ref{tab:cp}) compared to the target-only estimator.

\begin{table}[H]
\begin{subtable}[c]{0.5\textwidth}
\centering
\tiny
    \begin{tabular}{|l|*{5}{r|}}\hline
\backslashbox{$m_1$}{$m_2$} & 1 & 2 & 3 & 4 & 5 \\ \hline
1 & / & 0.021 & $-$0.128 & 0.021 & $-$0.287 \\ 
2 & 0.169 & / & 0.029 & $-$0.012 & $-$0.299 \\ 
3 & 0.219 & $-$0.003 & / & 0.109 & $-$0.067 \\ 
4 & 0.028 & $-$0.143 & $-$0.173 & / & $-$0.197 \\ 
5 & 0.436 & 0.387 & 0.086 & 0.124 & / \\ 
\hline
    \end{tabular}
\subcaption{Proposed estimator}
\end{subtable}
\begin{subtable}[c]{0.5\textwidth}
\centering
\tiny
 \begin{tabular}{|l|*{5}{r|}}\hline
\backslashbox{$m_1$}{$m_2$} & 1 & 2 & 3 & 4 & 5 \\ \hline
1 & / & 0.016 & 0.274 & 0.228 & 0.015 \\ 
2 & 0.261 & / & 0.272 & 0.099 & $-$0.097 \\ 
3 & 0.305 & 0.036 & / & 0.33 & 0.282 \\ 
4 & 0.18 & $-$0.15 & 0.094 & / & 0.032 \\ 
5 & 0.464 & 0.146 & 0.371 & 0.208 & / \\ 
\hline
    \end{tabular}
\subcaption{Target-only estimator}
\end{subtable}
\caption{Bias of estimators over 1000 replicates. The reported values are the original results multiplied by 100.}
\label{tab:bias}
\end{table}

\begin{table}[H]
\begin{subtable}[c]{0.5\textwidth}
\centering
\footnotesize
    \begin{tabular}{|l|*{5}{r|}}\hline
\backslashbox{$m_1$}{$m_2$} & 1 & 2 & 3 & 4 & 5 \\ \hline
1 & / & 0.063 & 0.063 & 0.064 & 0.066 \\ 
2 & 0.062 & / & 0.074 & 0.065 & 0.069 \\ 
3 & 0.066 & 0.078 & / & 0.070 & 0.073 \\ 
4 & 0.063 & 0.066 & 0.068 & / & 0.069 \\ 
5 & 0.069 & 0.076 & 0.074 & 0.073 & / \\ 
\hline
    \end{tabular}
\subcaption{Proposed estimator}
\end{subtable}
\begin{subtable}[c]{0.5\textwidth}
\centering
\footnotesize
 \begin{tabular}{|l|*{5}{r|}}\hline
\backslashbox{$m_1$}{$m_2$} & 1 & 2 & 3 & 4 & 5 \\ \hline
1 & / & 0.066 & 0.067 & 0.068 & 0.068 \\ 
2 & 0.064 & / & 0.076 & 0.068 & 0.072 \\ 
3 & 0.068 & 0.082 & / & 0.073 & 0.076 \\ 
4 & 0.067 & 0.070 & 0.071 & / & 0.071 \\ 
5 & 0.072 & 0.078 & 0.079 & 0.076 & / \\ 
\hline
    \end{tabular}
\subcaption{Target-only estimator}
\end{subtable}
\caption{RMSE of estimators over 1000 replicates.}
\label{tab:rmse}
\end{table}

\begin{table}[H]
\begin{subtable}[c]{0.5\textwidth}
\centering
    \begin{tabular}{|l|*{5}{r|}}\hline
\backslashbox{$m_1$}{$m_2$} & 1 & 2 & 3 & 4 & 5 \\ \hline
1 & / & 0.94 & 0.94 & 0.93 & 0.93 \\ 
2 & 0.95 & / & 0.93 & 0.93 & 0.94 \\ 
3 & 0.93 & 0.94 & / & 0.92 & 0.93 \\ 
4 & 0.93 & 0.93 & 0.92 & / & 0.92 \\ 
5 & 0.93 & 0.93 & 0.92 & 0.93 & / \\ 
\hline
    \end{tabular}
\subcaption{Proposed estimator}
\end{subtable}
\begin{subtable}[c]{0.5\textwidth}
\centering
 \begin{tabular}{|l|*{5}{r|}}\hline
\backslashbox{$m_1$}{$m_2$} & 1 & 2 & 3 & 4 & 5 \\ \hline
1 & / & 0.92 & 0.90 & 0.91 & 0.92 \\ 
2 & 0.93 & / & 0.91 & 0.91 & 0.92 \\ 
3 & 0.92 & 0.89 & / & 0.90 & 0.90 \\ 
4 & 0.91 & 0.91 & 0.90 & / & 0.91 \\ 
5 & 0.89 & 0.89 & 0.88 & 0.90 & / \\ 
\hline
    \end{tabular}
\subcaption{Target-only estimator}
\end{subtable}
\caption{Coverage Probability (CP) of the 95\% CI of estimators over 1000 replicates.}
\label{tab:cp}
\end{table}

\section{Results}

\subsection{Data Source}
The Congenital Heart Surgery Database (CHSD) was established by the Society of Thoracic Surgeons (STS) in 1994 to support quality improvement and safety efforts in pediatric and congenital cardiothoracic surgery. The database includes congenital heart surgery centers from North America and contains audited preoperative, intraoperative, and postoperative data. Data quality is enhanced through standard definitions, abstractor training, and random site audits. Data collection and quality of the CHSD have been described elsewhere \citep{overman2019ten}.  Briefly, data are sent to the STS data warehouse, where they are assembled, cleaned, and analyzed. Data quality checks include identification of unusual data values and inconsistencies across data fields, and an annual random audit of 10\% of participating centers to validate submitted data \citep{overman2019ten}. In these audits, all mortalities are reviewed by a surgeon auditor, and hospital surgical logs are compared with the case list submitted to the warehouse to ensure that all eligible cases are reported. 

Patients can have multiple admissions to the same surgical center. During an admission, patients may have multiple operative encounters, and within an operative encounter, they may experience multiple component procedures. We identified all cardiovascular operative encounters between January 1, 2016 and June 30, 2022 in U.S. surgical centers, and excluded patients based on standard criteria \citep{overman2019ten}. The index operative encounter during an admission within a hospital using the first (chronologically) operative encounter was identified and used as our unit of observation. Observations for patients having another surgical procedure within 30 days from the index operative encounter were merged to form a single episode of care and assigned to the surgical center responsible for the index encounter. Patients having two or more consecutive admissions and patients readmitted from another acute care or long-term care facility (without being discharged home or residing in a long-term care facility for 183 consecutive postoperative days) were also merged. Among the approximately 135,000 operative encounters, 3\% were Norwood procedures.

\subsection{Outcomes and Confounders}
We examine two outcomes: (i) operative mortality, defined by the STS-CHSD as death from any cause that occurred within 30 days after the index operative encounter, death at hospital discharge for patients alive at 30 days but still hospitalized in an acute care facility, or death within 183 days if discharged to a chronic care facility \citep{overman2013report}; and (ii) operative morbidity, defined as the occurrence in an encounter of any of six major complications: renal failure requiring dialysis, neurologic deficit persisting at discharge, arrhythmia requiring a permanent pacemaker, paralyzed diaphragm/phrenic nerve injury, mechanical circulatory support, or unplanned re-intervention \citep{pasquali2019development, o2019development}. 

\begin{figure}[H]
    \centering
    \includegraphics[width=0.85\textwidth]{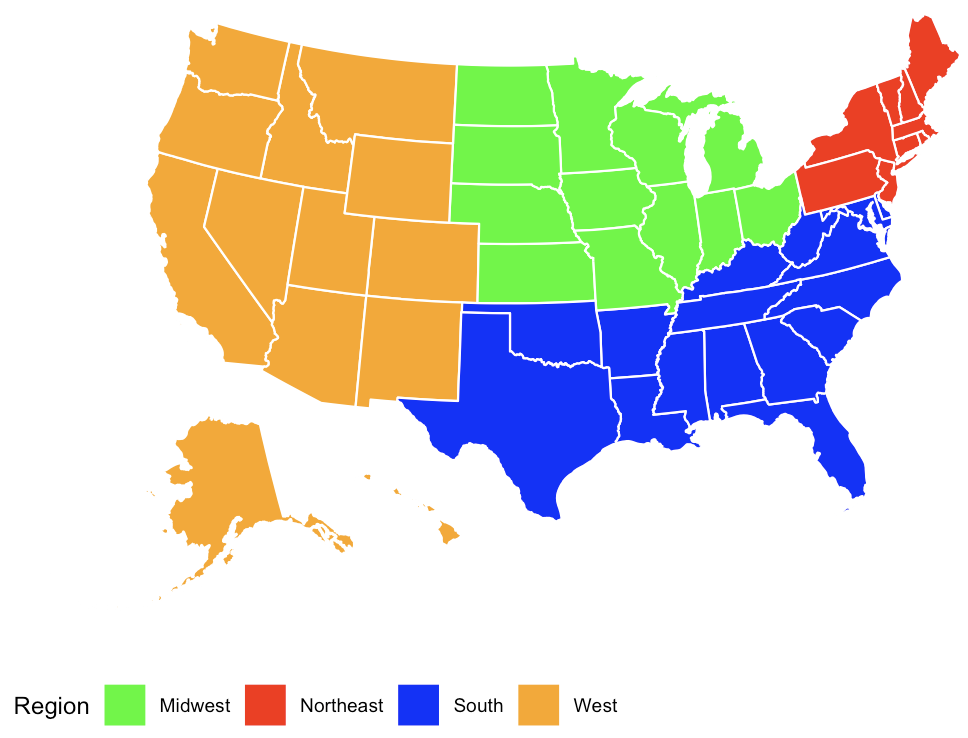}
    \caption{U.S. Census Bureau's geographic regions: (i) Northeast: Maine, New Hampshire, Vermont, Massachusetts, Connecticut, Rhode Island, New York, Pennsylvania, New Jersey, (ii) Midwest: North Dakota, South Dakota, Nebraska, Kansas, Minnesota, Iowa, Missouri, Wisconsin, Illinois, Indiana, Michigan, Ohio, (iii) South: Texas, Oklahoma, Arkansas, Louisiana, Mississippi, Alabama, Tennessee, Kentucky, Florida, Georgia, South Carolina, North Carolina, Virginia, West Virginia, District of Columbia, Maryland, Delaware,  (iv) West: Washington, Oregon, California, Nevada, Montana, Idaho, Wyoming, Utah, Arizona, Colorado, New Mexico, Alaska, Hawaii. }
    \label{fig:map}
\end{figure}

Race and ethnicity information was reported by the patient's guardian and categorized into three mutually exclusive categories: non-Hispanic Black, non-Hispanic Caucasian, and Hispanic. American Indian, Native Alaskan, Asian, and encounters missing race or ethnicity information are not included in the analysis due to limited samples. The geographic region of the surgical center was documented and categorized into four mutually exclusive regions according to the U.S. Census Bureau's geographic regions: (i) Northeast, (ii) Midwest,  (iii) South, and (iv) West \citep{census2020}. A map of the states to geographic regions is provided in Figure \ref{fig:map}. Four regions rather than nine divisions were selected to ensure that there were at least 10 surgical centers in each region. 

Confounders that may relate to outcomes included those reported previously spanning general demographic factors, genetic syndromes, chromosomal abnormalities, non-cardiac anomalies, pre-operative factors, and a variety of Norwood procedure-specific factors found in the STS-CHSD \citep{tabbutt2012risk}. Pre-operative factors included cardiopulmonary resuscitation, preoperative complete AV block, mechanical circulatory support, shock, mechanical ventilation, renal dysfunction, neurological deficit, seizure, stroke, CVA, or intracranial hemorrhage, sepsis, necrotizing enterocolitis, coagulation disorder, respiratory syncytial virus infection, and tracheostomy. Norwood procedure-specific factors included the following factors: an ascending aorta diameter of $<2$ mm, aortic valve atresia, aortic valve stenosis, mitral valve atresia, mitral valve stenosis, left ventricle to coronary artery sinusoids, intact atrial septum, obstructed pulmonary venous return with restrictive atrial septic defect (ASD), atrioventricular valve regurgitation (AR) grades 3 and 4, aberrant right subclavian artery origin, and right vs. left ventricular dominance. We included a total of 32 confounders.

\subsection{Cohort characteristics and unadjusted outcomes}

We identified $3{,}588$ index Norwood operative encounters from $100$ surgical centers. Of these, $1{,}457$ $(40.6\%)$ were female, $329$ $(9.2\%)$ were born premature ($< 37$ weeks), $707$ $(19.7\%)$ had at least one chromosomal abnormality, and $650$ $(18.1\%)$ had at least one syndromic abnormality. The primary diagnosis was HLHS in $2{,}622$ $(73.1\%)$. The racial/ethnic category was recorded as $649$ $(18.1\%)$ for non-Hispanic Black, $758$ (21.1\%) for Hispanic, and $2{,}181$ $(60.8\%)$ for non-Hispanic Caucasian. There were $19$, $14$, $28$, and $39$ surgical centers in the South, Midwest, West, and Northeast regions, with operative encounters totaling $790$ $(22.0\%)$, $536$ $(14.9\%)$, $998$ $(27.8\%)$, and $1{,}264$ $(35.2\%)$, respectively. Other baseline characteristics are provided in Table \ref{tab:1}.  

\begin{table}
    \scriptsize
    \centering
    \caption{Baseline characteristics of patients undergoing the Norwood procedure.}
\begin{tabular}[h]{l|llll|l}
\toprule
  & South & Midwest & West & Northeast & Overall\\
\midrule
 & ($n_1=790$) & ($n_2=536$) & ($n_3=998$) & ($n_4=1264$) & ($N=3588$)\\
Race/ethnicity &  &  &  &  & \\
\quad Caucasian (non-Hispanic) & 419 (53.0\%) & 350 (65.3\%) & 574 (57.5\%) & 672 (53.2\%) & 2015 (56.2\%)\\
\quad Hispanic & 202 (25.6\%) & 71 (13.2\%) & 277 (27.8\%) & 208 (16.5\%) & 758 (21.1\%)\\
\quad Black (non-Hispanic) & 153 (19.4\%) & 100 (18.7\%) & 86 (8.6\%) & 310 (24.5\%) & 649 (18.1\%)\\
\quad Asian (non-Hispanic) & 8 (1.0\%) & 6 (1.1\%) & 25 (2.5\%) & 31 (2.5\%) & 70 (2.0\%)\\
\quad Native Am./Pac. Islander & 6 (0.8\%) & 1 (0.2\%) & 11 (1.1\%) & 1 (0.1\%) & 19 (0.5\%)\\
\quad Other (non-Hispanic) & 2 (0.3\%) & 8 (1.5\%) & 25 (2.5\%) & 42 (3.3\%) & 77 (2.1\%)\\
\midrule
Year of Norwood &  &  &  &  & \\
\quad 2016 & 125 (15.8\%) & 84 (15.7\%) & 177 (17.7\%) & 191 (15.1\%) & 577 (16.1\%)\\
\quad 2017 & 126 (15.9\%) & 96 (17.9\%) & 165 (16.5\%) & 204 (16.1\%) & 591 (16.5\%)\\
\quad 2018 & 142 (18.0\%) & 91 (17.0\%) & 150 (15.0\%) & 187 (14.8\%) & 570 (15.9\%)\\
\quad 2019 & 128 (16.2\%) & 92 (17.2\%) & 170 (17.0\%) & 203 (16.1\%) & 593 (16.5\%)\\
\quad 2020 & 116 (14.7\%) & 69 (12.9\%) & 136 (13.6\%) & 221 (17.5\%) & 542 (15.1\%)\\
\quad 2021 & 97 (12.3\%) & 66 (12.3\%) & 132 (13.2\%) & 191 (15.1\%) & 486 (13.5\%)\\
\quad 2022 (through June) & 56 (7.1\%) & 38 (7.1\%) & 68 (6.8\%) & 67 (5.3\%) & 229 (6.4\%)\\
\midrule
Age in days: Mean (SD) 
& 6.9 (6.0) & 6.7 (5.6) & 7.1 (6.5) & 6.8 (6.7) & 6.9 (6.3)\\
\midrule
Sex &  &  &  &  & \\
\quad Female & 318 (40.3\%) & 203 (37.9\%) & 435 (43.6\%) & 501 (39.6\%) & 1457 (40.6\%)\\
\quad Male & 472 (59.7\%) & 333 (62.1\%) & 563 (56.4\%) & 763 (60.4\%) & 2131 (59.4\%)\\
\midrule
Birthweight in kg: Mean (SD) & 3.25 (0.51) & 3.29 (0.49) & 3.24 (0.49) & 3.22 (0.53) & 3.24 (0.51)\\
\midrule
Birth height in cm: Mean (SD) & 49.6 (4.71) & 49.9 (5.53) & 49.5 (3.14) & 49.1 (3.21) & 49.4 (3.98)\\
\midrule
Premature &  &  &  &  & \\
\quad $>37$ weeks & 686 (86.8\%) & 504 (94.0\%) & 911 (91.3\%) & 1157 (91.5\%) & 3258 (90.8\%)\\
\quad $<37$ weeks & 103 (13.0\%) & 32 (6.0\%) & 87 (8.7\%) & 107 (8.5\%) & 329 (9.2\%)\\
\quad Missing & 1 (0.1\%) & 0 (0\%) & 0 (0\%) & 0 (0\%) & 1 (0.0\%)\\
\midrule
Reoperation within admission &  &  &  &  & \\
\quad Not & 780 (98.7\%) & 506 (94.4\%) & 981 (98.3\%) & 1234 (97.6\%) & 3501 (97.6\%)\\
\quad Planned  & 10 (1.3\%) & 18 (3.4\%) & 13 (1.3\%) & 22 (1.7\%) & 63 (1.8\%)\\
\quad Unplanned  & 0 (0\%) & 12 (2.2\%) & 4 (0.4\%) & 8 (0.6\%) & 24 (0.7\%)\\
\midrule
Cardiac anomaly &  &  &  &  & \\
\quad Antenatally diagnosed & 643 (81.4\%) & 462 (86.2\%) & 838 (84.0\%) & 1104 (87.3\%) & 3047 (84.9\%)\\
\quad Not antenatally diagnosed & 147 (18.6\%) & 74 (13.8\%) & 157 (15.7\%) & 159 (12.6\%) & 537 (15.0\%)\\
\quad Missing & 0 (0\%) & 0 (0\%) & 3 (0.3\%) & 1 (0.1\%) & 4 (0.1\%)\\
\midrule
Non-cardiac abnormality &  &  &  &  & \\
\quad One or more & 153 (19.4\%) & 124 (23.1\%) & 181 (18.1\%) & 254 (20.1\%) & 712 (19.8\%)\\
\quad None & 637 (80.6\%) & 412 (76.9\%) & 817 (81.9\%) & 1010 (79.9\%) & 2876 (80.2\%)\\
\midrule
Chromosomal abnormality &  &  &  &  & \\
\quad One or more & 169 (21.4\%) & 122 (22.8\%) & 186 (18.6\%) & 230 (18.2\%) & 707 (19.7\%)\\
\quad None & 621 (78.6\%) & 412 (76.9\%) & 812 (81.4\%) & 1034 (81.8\%) & 2879 (80.2\%)\\
\quad Missing & 0 (0\%) & 2 (0.4\%) & 0 (0\%) & 0 (0\%) & 2 (0.1\%)\\
\midrule
Syndromic abnormality &  &  &  &  & \\
\quad One or more & 162 (20.5\%) & 104 (19.4\%) & 173 (17.3\%) & 211 (16.7\%) & 650 (18.1\%)\\
\quad None & 628 (79.5\%) & 432 (80.6\%) & 825 (82.7\%) & 1053 (83.3\%) & 2938 (81.9\%)\\
\midrule
Pre-operative factors &  &  &  &  & \\
\quad One or more & 453 (57.3\%) & 266 (49.6\%) & 490 (49.1\%) & 717 (56.7\%) & 1926 (53.7\%)\\
\quad None & 337 (42.7\%) & 270 (50.4\%) & 508 (50.9\%) & 547 (43.3\%) & 1662 (46.3\%)\\
\midrule
Fundamental diagnosis &  &  &  &  & \\
\quad HLHS & 606 (76.7\%) & 383 (71.5\%) & 710 (71.1\%) & 917 (72.5\%) & 2622 (73.1\%)\\
\quad DILV & 38 (4.8\%) & 36 (6.7\%) & 63 (6.3\%) & 56 (4.4\%) & 193 (5.4\%)\\
\quad Unbalanced AV canal & 19 (2.4\%) & 25 (4.7\%) & 51 (5.1\%) & 58 (4.6\%) & 153 (4.3\%)\\
\quad Tricuspid atresia & 21 (2.7\%) & 18 (3.4\%) & 41 (4.1\%) & 50 (4.0\%) & 130 (3.6\%)\\
\quad Mitral atresia & 22 (2.8\%) & 13 (2.4\%) & 25 (2.5\%) & 29 (2.3\%) & 89 (2.5\%)\\
\quad Other & 84 (10.6\%) & 61 (11.4\%) & 108 (10.8\%) & 148 (11.7\%) & 401 (11.2\%)\\
\addlinespace
\bottomrule
\end{tabular}
\label{tab:1}
\end{table}

The overall unadjusted mortality (morbidity) rates were $11.8\%$ ($25.9\%$) and varied across racial and ethnic groups: non-Hispanic Caucasian $11.1\%$ ($26.1\%$), Hispanic patients $11.6\%$ ($22.6\%$), non-Hispanic Black patients $12.9\%$ ($28.0\%$), non-Hispanic Asian patients $17.1\%$ ($33.8\%$). Regional variation was less: Northeast $11.1\%$ ($25.6\%$), South $11.5\%$ ($25.9\%)$, Midwest $11.9\%$ ($28.7\%$), West $12.7\%$ ($24.8\%$). 

\begin{figure}[H]
    \centering
    \includegraphics[width=0.75\textwidth]{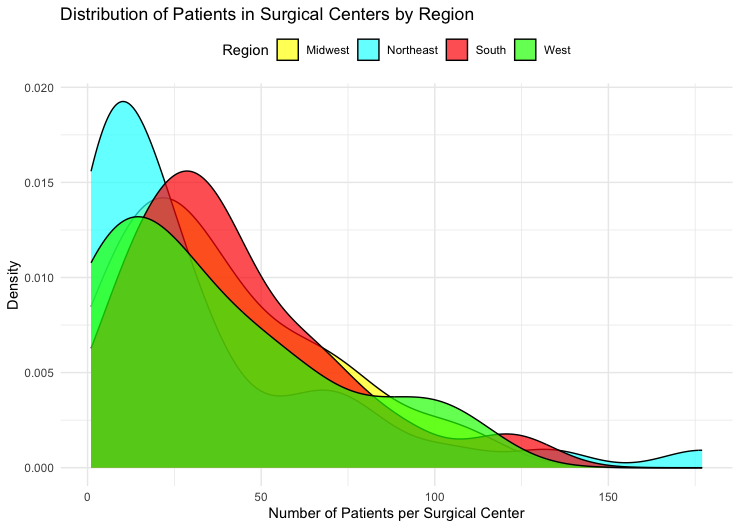}
    \caption{Distribution of Norwood procedures per surgical center in each of the geographic regions. Of the 100 total hospitals, the Midwest consists of 14 hospitals, the Northeast 39 hospitals, the South 19 hospitals, and the West 28 hospitals.}
    \label{fig:volume}
\end{figure}

Given the previously reported associations of center volume with outcome, we also evaluated the distribution of surgical center volume across regions. Center volume varied by region, with institutional volumes being generally higher in the South and overall lower in the Northeast, although the Northeast also had some of the individual highest-volume centers (Figure \ref{fig:volume}). 

\subsection{Causal effect estimates by race and ethnicity across geographic regions}

\begin{figure}[H]
    \centering
    \includegraphics[width=\textwidth]{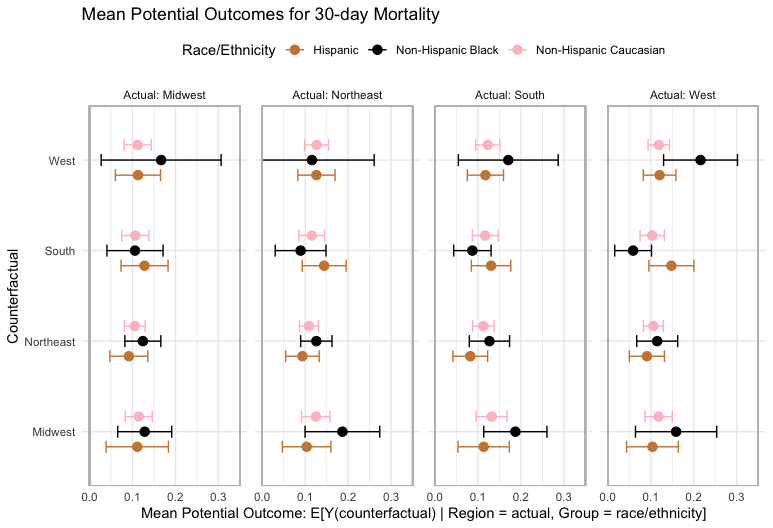}
    \caption{Estimates of the mean potential outcome for $30$-day mortality. Each of the $16$ panels represents estimated $30$-day mortality for a target population  had they undergone the Norwood procedure in another (potentially counterfactual) region. Rows represent counterfactual regions. Columns represent actual regions. Black dots and lines represent non-Hispanic black patients, pink represents non-Hispanic Caucasian patients, and brown represents Hispanic patients. }
    \label{fig:mpo.mort}
\end{figure}

\begin{figure}[H]
    \centering
    \includegraphics[width=\textwidth]{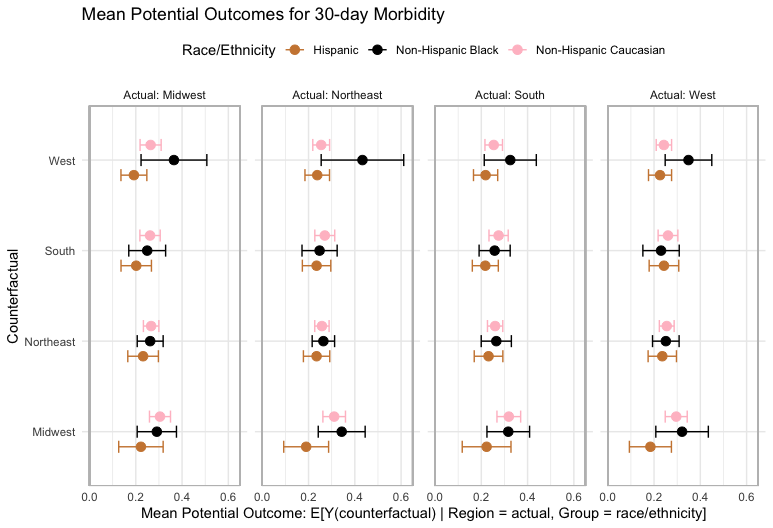}
    \caption{Estimates of the mean potential outcome for $30$-day morbidity. Each of the $16$ panels represents estimated $30$-day morbidity for a target population  had they undergone the Norwood procedure in another (potentially counterfactual) region. Rows represent the different counterfactual regions. Columns represent actual regions. Black dots and lines represent non-Hispanic Black patients, pink represents non-Hispanic Caucasian patients, and brown represents Hispanic patients. }
    \label{fig:mpo.morb}
\end{figure}

Our causal models estimated that non-Hispanic Black patients who underwent the Norwood procedure in the South region (estimate = 8.7\% [95\% CI: 4.4\%, 13.0\%] would have fared worse had they been hypothetically treated in another region, and Non-Hispanic Black patients who underwent the Norwood procedure in the West region (estimate = 21.6\% [95\% CI: 13.0\%, 30.2\%] would have fared better had they been hypothetically treated in another region (Figure \ref{fig:mpo.mort}). For non-Hispanic Caucasian patients, the risk of mortality tended to be consistent across regions and lower than that for non-Hispanic Black patients (except for the South region). Similar patterns were observed for morbidity (Figure \ref{fig:mpo.morb}).

\begin{figure}[H]
    \centering
    \includegraphics[width=\textwidth]{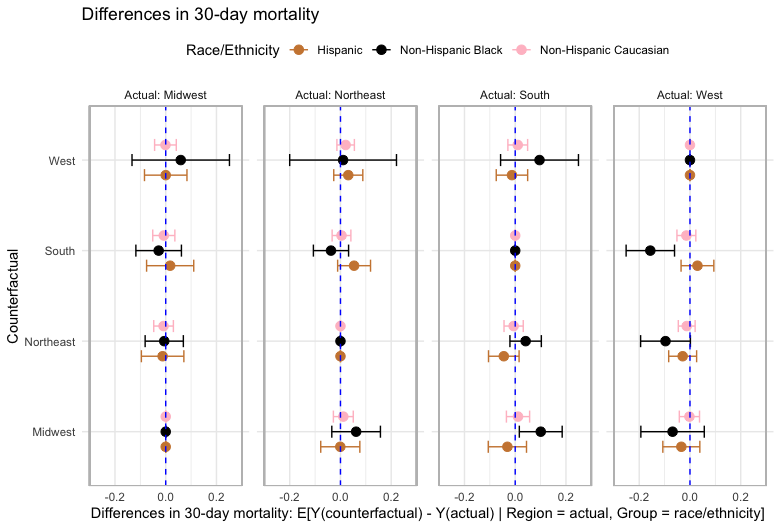}
    \caption{Estimates of the average region effect for $30$-day mortality. Each of the $16$ panels represents estimated differences in $30$-day mortality for a target population had they undergone the Norwood procedure in another (potentially counterfactual) region. Values to the right (left) of the vertical blue dashed line indicate lower mortality for patients in actual (counterfactual) region Rows represent counterfactual regions. Columns represent actual regions. Black dots and lines represent non-Hispanic black patients, pink represents non-Hispanic Caucasian patients, and brown represents Hispanic patients. }
    \label{fig:att.mort}
\end{figure}

\begin{figure}[H]
    \centering
    \includegraphics[width=\textwidth]{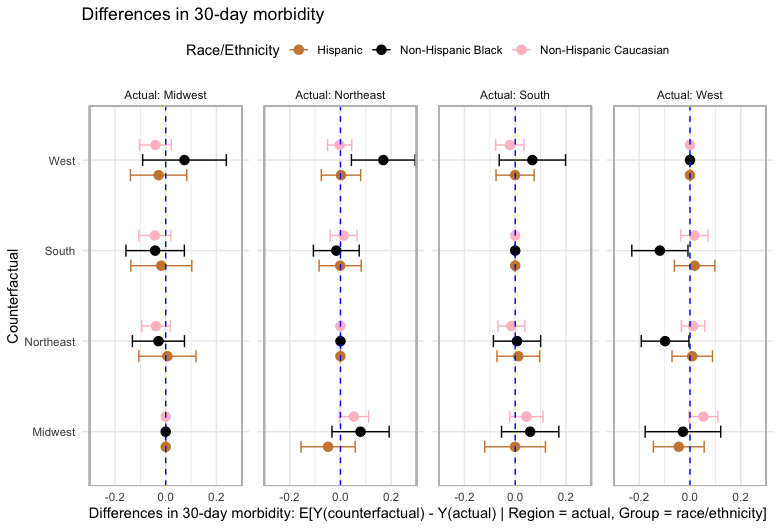}
    \caption{Estimates of the average region effect for $30$-day morbidity. Each of the $16$ panels represents estimated differences in $30$-day morbidity for a target population had they undergone the Norwood procedure in another (potentially counterfactual) region. {Values to the right (left) of the vertical blue dashed line indicate lower mortality for patients in actual (counterfactual) region.} Rows represent counterfactual regions. Columns represent actual regions. Values to the right (left) of the vertical dashed line indicate lower morbidity for patients in the target (counterfactual) region.  Black dots and lines represent non-Hispanic Black patients, pink represents non-Hispanic Caucasian patients, and brown represents Hispanic patients.CIs not adjusted for multiplicity.}
    \label{fig:att.morb}
\end{figure}

We also found larger variability in outcomes for non-Hispanic Black and Hispanic patients compared to non-Hispanic Caucasian patients, with the largest variability among non-Hispanic Black patients for both mortality (Figure \ref{fig:att.mort}) and morbidity outcomes (Figure \ref{fig:att.morb}). For non-Hispanic Black patients, regional effects differed significantly. For instance, for non-Hispanic Black patients undergoing the Norwood procedure in the South region of the U.S., had they instead undergone their surgery in the Midwest region, their mortality risk would be expected to be higher (absolute difference 10.1\% [95\% CI: 1.6\%, 18.5\%]). For non-Hispanic Caucasian patients, the estimated risk of mortality does not vary across regions. Findings for Hispanic patients varied more than for non-Hispanic Caucasian patients but less than for non-Hispanic Black patients. 

\section{Sensitivity Analysis}

\begin{figure}[H]
    \centering
    \includegraphics[width=\textwidth]{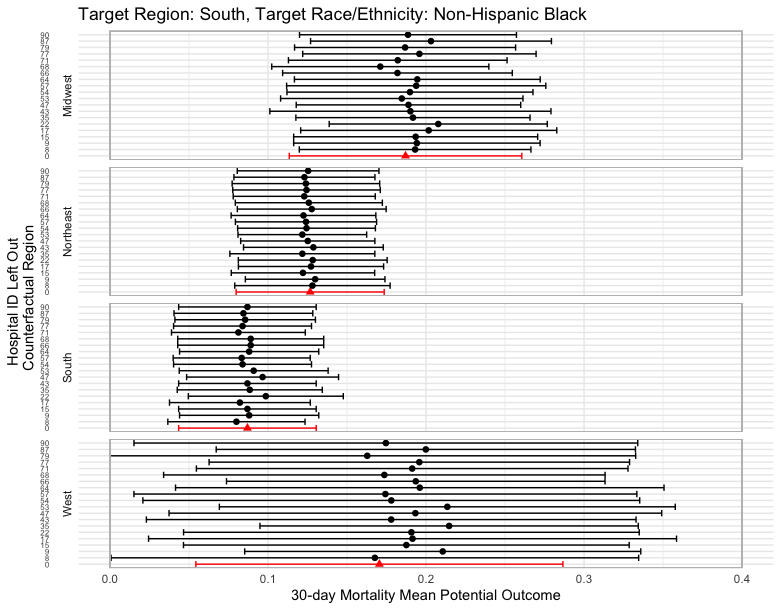}
    \caption{Leave-one-out surgical center sensitivity analysis of the mean potential outcome for $30$-day mortality. The target population is defined as non-Hispanic black patients who underwent the Norwood procedure in the South region. The region consists of $19$ surgical centers. Each of the $19$ sensitivity analyses excludes one surgical center, i.e., each row represents the counterfactual mean potential outcome when the corresponding hospital ID is left out. The complete case analysis is given as the last line in each counterfactual grouping in red with a triangle point estimate.}
    \label{fig:loo.south}
\end{figure}

\begin{figure}[H]
    \centering
    \includegraphics[width=\textwidth]{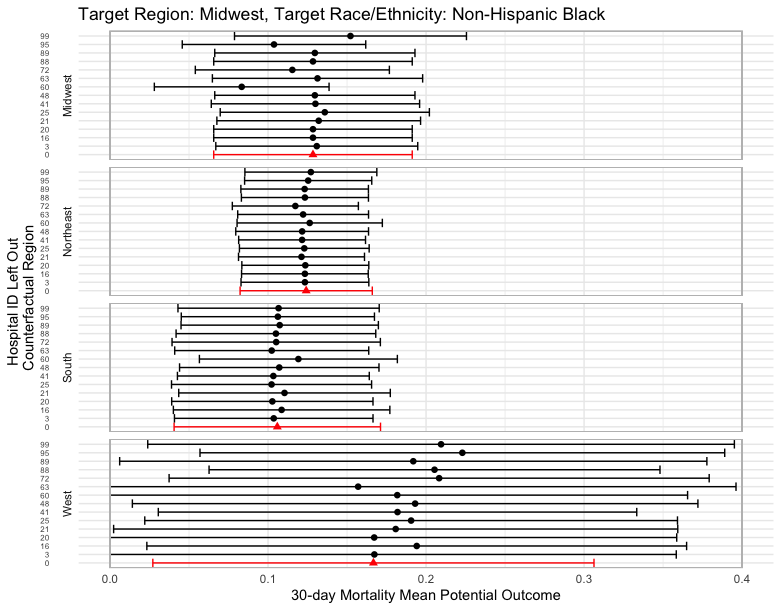}
    \caption{Leave-one-out surgical center sensitivity analysis of the mean potential outcome for $30$-day mortality. The target population is defined as non-Hispanic Black patients who underwent the Norwood procedure in the Midwest region. The region consists of $14$ surgical centers. Each of the $14$ sensitivity analyses excludes one surgical center, i.e., each row represents the counterfactual mean potential outcome when the corresponding hospital ID is left out. The complete case analysis is given as the last line in each counterfactual grouping in red with a triangle point estimate.}
    \label{fig:loo.midwest}
\end{figure}

\begin{figure}[H]
    \centering
    \includegraphics[width=\textwidth]{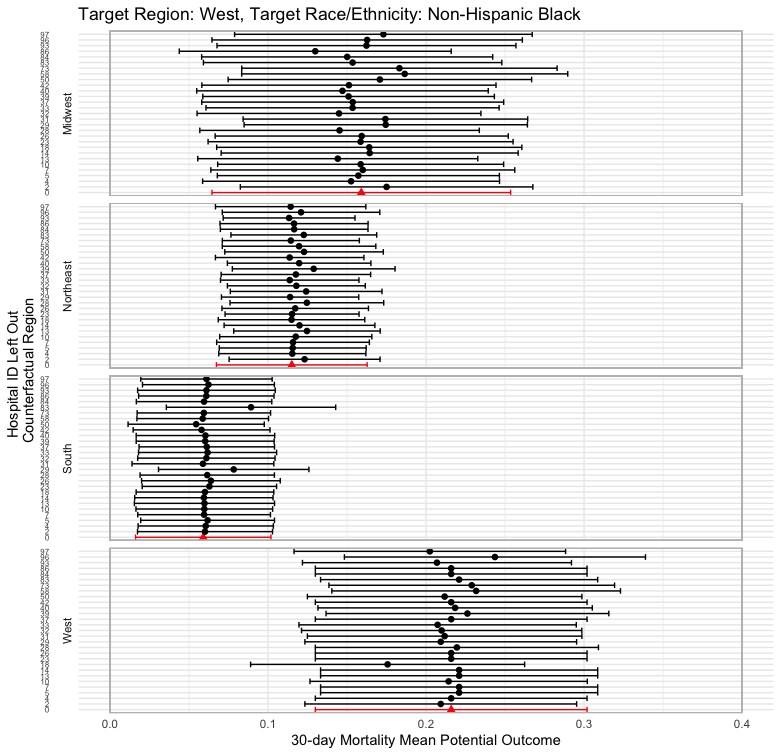}
    \caption{Leave-one-out surgical center sensitivity analysis of the mean potential outcome for $30$-day mortality. The target population is defined as non-Hispanic Black patients who underwent the Norwood procedure in the West region. The region consists of $28$ surgical centers. Each of the $28$ sensitivity analyses excludes one surgical center, i.e., each row represents the counterfactual mean potential outcome when the corresponding hospital ID is left out. The complete case analysis is given as the last line in each counterfactual grouping in red with a triangle point estimate.}
    \label{fig:loo.west}
\end{figure}

\begin{figure}[H]
    \centering
    \includegraphics[width=\textwidth]{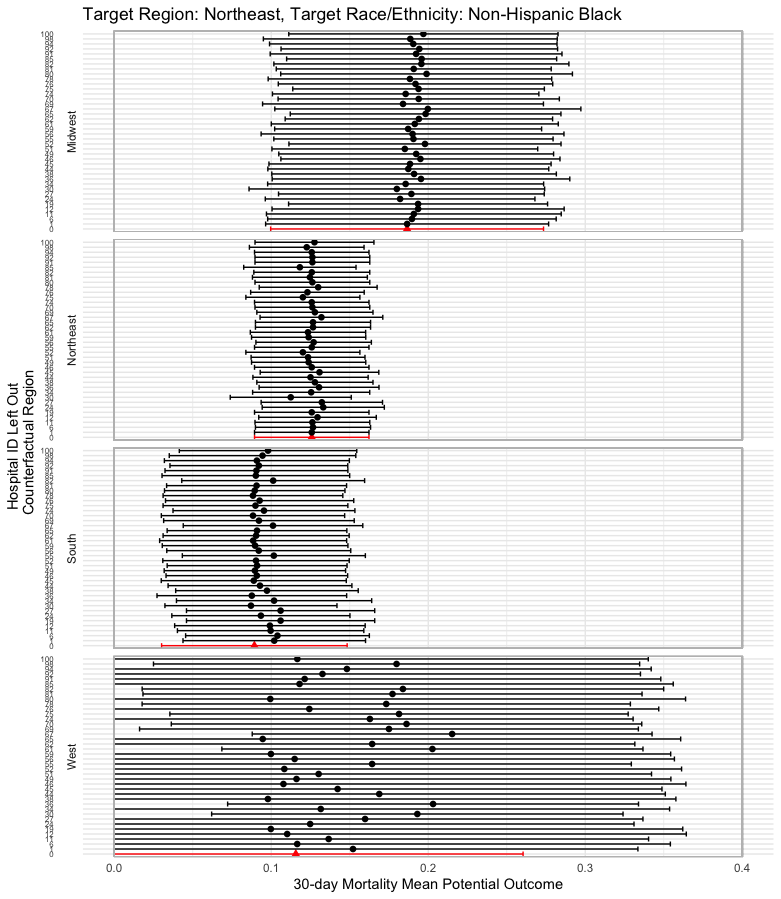}
    \caption{Leave-one-out surgical center sensitivity analysis of the mean potential outcome for $30$-day mortality. The target population is defined as non-Hispanic Black patients who underwent the Norwood procedure in the Northeast region. The region consists of $39$ surgical centers. Each of the $39$ sensitivity analyses excludes one surgical center, i.e., each row represents the counterfactual mean potential outcome when the corresponding hospital ID is left out. The analysis including all centers in the region is given as the last line in each counterfactual grouping in red with a triangle point estimate.}
    \label{fig:loo.northeast}
\end{figure}

To examine whether the heterogeneity in regional effects for non-Hispanic Black patients was driven by specific surgical centers, we conducted a leave-one-center-out sensitivity analysis. For instance, there are $19$ surgical centers in the South region. When the target region is the South region, we repeated the analyses for the outcomes $19$ times, in each analysis removing one of the surgical centers located in the South leaving $18$ of the surgical centers contributing data (Figure \ref{fig:loo.south}).  Sensitivity analyses for the other regions are reported when the target region is the Midwest region ($14$ hospitals) (Figure \ref{fig:loo.midwest}), the West region ($28$ hospitals) (Figure \ref{fig:loo.west}), and the Northeast region ($39$ hospitals) (Figure \ref{fig:loo.northeast}). Our findings are robust across centers, suggesting no single center is driving the results.

\newpage

\section{Discussion}\label{sec3}

We observed racial and ethnic differences across geographic regions in operative mortality and morbidity outcomes among patients with CHD undergoing the Norwood procedure. These differences persisted in our causal models after adjusting for detailed clinical risk factors collected using uniform definitions and auditing. We found that post-surgical outcomes differed across geographic regions specifically for non-Hispanic Black patients but did not for non-Hispanic Caucasian patients.

We adopted a novel transfer learning causal approach that improves upon existing methods that are unable to accommodate analyses of underrepresented groups due to the small sample sizes and relatively sparse number of cases within risk factor groups. Among two currently available methods, one approach completely pools data and uses regression modeling with indicators for racial and ethnic groups as well as for geographic regions. Often the relationship between risk factors and outcomes does not vary across subgroups.  In the other approach,  each racial and ethnic group is analyzed separately, but this is limited by small sample sizes.  In our approach, we account for multiple differences between target and source populations, including differences in risk factor distributions (e.g., some racial and ethnic populations may have more chromosomal abnormalities), in the relationship between risk factors and region where surgery was performed (e.g., patients with the same covariate profile may have different probabilities of undergoing surgery in a specific region), and in the relationship of risk factors with outcomes (e.g., patients with the same covariate profile may experience different outcomes). Our approach makes no a priori assumptions on a common relationship of risk factors and outcomes between specific race/ethnicity groups. Instead, for greater flexibility, we estimate parameters data-adaptively via transfer learning.

In sensitivity analyses, the inclusion or exclusion of no specific surgical center changed our results. However, we did find center volume variability in the number of Norwood procedures across regions and this could, in part, account for some of the differences in patient outcomes observed. Fewer non-Hispanic Black patients underwent a Norwood procedure in the West region and our analyses showed that non-Hispanic Black patients were expected to experience lower mortality had they been treated in any of the other regions. Non-Hispanic Black patients treated in the South region where center volumes were generally higher in our analysis (Fig \ref{fig:volume}), were expected to experience the lowest mortality across regions, consistent with prior studies reporting an association between higher volume and improved post-operative outcomes \citep{maruthappu2015influence, williamson2022center}. Racial and ethnic differences in access to high-quality surgical centers in different geographic regions, including travel across geographic region borders, may also account for observed differences in outcomes. Previous studies have shown that non-Hispanic Black patients with coronary heart disease are more likely to seek care at lower-quality cardiac centers, even when higher-quality hospitals are geographically closer to their homes \citep{popescu2010racial, popescu2019contributions, udine2021geographical, sengupta2023impact, bucholz2018neighborhood}.  However, our sensitivity analysis suggested no single center impacted our findings.

Our findings that differences in outcomes persist after causal modeling across geographic regions for non-Hispanic Black patients but not for non-Hispanic Caucasian patients suggest that further research is needed to understand the heterogeneity in outcomes, underlying factors, and potential interventions specifically for the non-Hispanic Black population.

As with all observational data analyses, our findings may be vulnerable to unmeasured confounding. While we included a rich set of confounders, we did not have access to variables that reflect in a more granular fashion the socioeconomic status, e.g., child opportunity index, variables that measure the severity of residual lesions, and an overall physiologic assessment of the preoperative status, as has been described previously \citep{berger2017morbidity}. We fit parametric outcome regression models and used penalized regression in the transfer learning step. Given the difficulty in correctly specifying regression models with high-dimensional covariates, we may have misspecified the regression models.  Future research using nonparametric machine-learning methods to estimate outcome regression models and transfer learning parameters would increase robustness.

\section*{Acknowledgments}
This work was supported, in part, by Grant HL5R01HL162893 from the National Heart, Lung, and Blood Institute from the US National Institutes of Health. The data analyzed in this study were provided to the investigators through The Society of Thoracic Surgeons' Task Force on Funded Research Program. 

\bibliographystyle{agsm}
\bibliography{ref}

\end{document}